\documentclass{elsart3}

\usepackage{graphicx}
\usepackage{subfigure}
\usepackage{latexsym}

\begin{document}

\newcommand{\greeksym}[1]{{\usefont{U}{psy}{m}{n}#1}}
\newcommand{\umu}{\mbox{\greeksym{m}}}
\newcommand{\udelta}{\mbox{\greeksym{d}}}
\newcommand{\uDelta}{\mbox{\greeksym{D}}}
\newcommand{\uOmega}{\mbox{\greeksym{W}}}
\newcommand{\uPi}{\mbox{\greeksym{P}}}
\newcommand{\ualpha}{\mbox{\greeksym{a}}}

\newcommand{\mrm}{\mathrm}
\newcommand{\Neq}{\mrm{n}_{\mrm{eq}}/\mrm{cm}^{2}}
\newcommand{\fns}{\footnotesize}
\newcommand{\scrs}{\scriptsize}

\sloppy

\begin{frontmatter}

\title{Performance of the CMS Pixel Detector at an upgraded LHC}
\author{R.\,Horisberger},
\author{D.\,Kotli\'nski}, and
\author{T.\,Rohe\thanksref{corr}}
\address{Paul Scherrer Institut, 5232 Villigen PSI, Switzerland}
\thanks[corr]{Corresponding author; e-mail: Tilman.Rohe@cern.ch}
\begin{abstract}
The CMS experiment will include a pixel detector for pattern recognition
and vertexing. It will consist of three barrel layers and two endcaps
on each side, providing three space-points up to a pseudorapidity of
2.1. Taking into account the expected limitations of its performance 
in the LHC environment an 8-9 layer pixel detector for an upgraded LHC
is discussed.

Key words: LHC, super LHC, CMS, tracking, pixel, silicon, radiation hardness
\end{abstract}
\end{frontmatter}

\section{Introduction}

The tracker of the CMS experiment consistes on silicon detectors only. The
region with a distance to the beam pipe between 22 and 115\,cm is equipped
with  10 layers of single sided silicon strip detectors covering an area 
of almost 200\,m$^2$ with about $10\times 10^{6}$ readout channels 
(strips)\cite{ref:tdr,ref:addendum}. 
The smaller radii will be equipped with a 1-m$^2$-pixel detector containing
about $60\times 10^{6}$ readout channels (pixels) providing three precision 
space points up to a pseudorapidity of 2.1~\cite{ref:tdr}. These unambiguous
space points allow an effective pattern recognition in the multiple track environment 
close to the LHC interaction point. Further the precision of the measurement is
used to identify displaced vertices for the tagging of b-jets and $\tau$-leptons.

In its final configuration the pixel detector will
consist of three barrel layers
and two end disks at each side. The barrels will be 53\,cm long and placed at
radii of 4.4\,cm, 7.3\,cm, and 10.2\,cm.
They cover an area of about $0.8\,\mrm{m}^{2}$ with
roughly 800 modules. The end disks are located at a mean distance from the
interaction point of 34.5\,cm and 46.5\,cm. The area of the 96 turbine blade
shaped modules in the disks sums up to about $0.28\,\mrm{m}^{2}$.

To achieve the necessary spatial resolution, analogue interpolation between neighbouring
channels will be performed. The 4-T-strong magnetic field in the CMS inner
detector causes a Lorentz angle of up to $26^{\circ}$ (unirradiated detector
at 100\,V bias voltage) \cite{ref:dorokhov-fi} and distributes the signal over
several pixels. For this reason the pixel size of $100\times 150\,\mrm{m}^{2}$
was chosen.

The two main challenges for the design of the pixel detector are the
high track rate and the high level of radiation. The former concerns the
architecture of the readout electronics. 
For the 4\,cm layer it has to locally store  the hit
information of about $20\times 10^{6}$ tracks per second and cm$^2$ 
at full LHC luminosity ($10^{34}\,\mrm{cm}^{-2}\mrm{s}^{-1}$)
for the latency of the 1$^{\mrm{st}}$ level trigger ($3.2\,\umu$s).
The high radiation level mainly affects the charge collection properties of the sensor,
which degrades steadily. In order to preserve a spatial resolution of better than
$20\,\umu$m, which is required for efficient b-tagging, the pixel modules will be replaced after a fluence of $6\times 10^{14}\,\Neq$.

A possible luminosity upgrade of LHC is currently being discussed. With a minor hardware upgrade
a luminosity  of $3-5\times 10^{34}\,$cm$^{-2}$s$^{-1}$ might be reached. Later major 
investments will aim for a luminosity of $10^{35}\,$cm$^{-2}$s$^{-1}$ \cite{ref:slhc}.
Already the first stage of the accelerator upgrade will require a major upgrade of
the tracker. The granularity of the strip detectors will become insufficient and 
these detectors will have to 
be replaced by pixel devices up to a radius of about 60\,cm. As the area to 
be covered by such a system is in the order of 10\,m$^{2}$, the choice of the 
detector concept will be governed by financial considerations. Ideas for
such cost effective pixel detectors are given in section~\ref{sec:mid-radii} 
and~\ref{sec:large-radii}. 

\begin{figure}
\centering\includegraphics[width=1\linewidth]{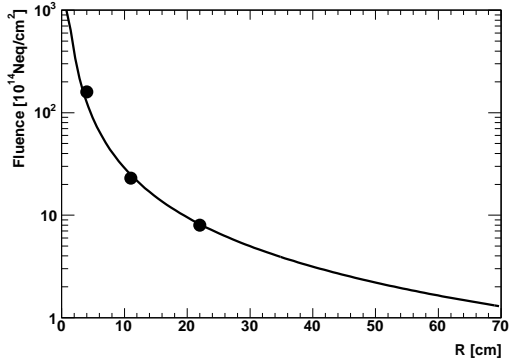}
\caption{Estimate of the radial dependence of the particle fluence
     for an integrated luminosity of 2500\,fb$^{-1}$\label{fig:slhc-fluence}}
\end{figure}

The inner regions of the tracker will have to face 
an unprecedented track rate 
and radiation level as shown in Fig.~\ref{fig:slhc-fluence}. 
The detectors placed at a radius of 4\,cm
have to withstand the presently unreached particle fluence of $\Phi\approx 10^{16}\,\Neq$
or must be replaced frequently. The operation limit of
a present type hybrid pixel system is discussed in section~\ref{sec:limits}. 
This is used as a starting point for a proposal of an 8-9 layer pixel system 
for an upgraded CMS pixel tracker.

\section{Limitations of a present-Type Pixel Detector at an upgraded LHC\label{sec:limits}}

The limiting factors for the operation of a hybrid pixel system at
an upgraded LHC remain the same as for the design of the initial 
CMS pixel detector: the track rate for the readout electronics
and the radiation induced degradation for the sensors.

\subsection{Sensor}

Since the late 1980's the radiation induced degradation of silicon detectors
has been systematically studied. The main effects are the increase of the
leakage current, trapping and the space charge. While the space
charge increase is reduced for silicon containing a high concentration
of oxygen \cite{ref:rose}, no dependence on the starting material could be
found for the other two parameters. Assuming that the increased leakage current
can be controlled by cooling, the reduction of the signal by
charge trapping presently sets the ultimate limit of the use of silicon 
detectors. 

Trapping of the signal charge is caused by irradiation induced energy
levels in the band gap. It can be described by the trapping
time which is inversely proportional to particle fluence.
\begin{figure}
\centering\includegraphics[width=1\linewidth]{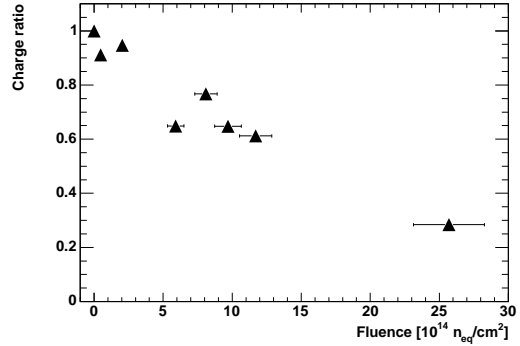}
\caption{Relative signal height obtained from a silicon pixel detector as a
     function of the irradiation fluence \cite{ref:rohe-fi}\label{fig:sig}}
\end{figure}
This is illustrated in Fig.~\ref{fig:sig} were the signal degradation
in a $285\,\umu$m thick pixel detector with increasing particle fluence is shown.
In this measurement the bias voltage was adjusted for each fluence to values
between 100\,V and 600\,V \cite{ref:rohe-fi}. A further increase will not
increase the signal level considerably. Up to a fluence of $\Phi\approx 10^{15}\,\Neq$
a total signal above 12\,000 electrons originating from the whole detector
thickness \cite{ref:dorokhov-fi} is achieved which is sufficient for an efficient
particle detection at comparator thresholds of 2000-3000 electrons.

At higher fluences the signal steadily decreases further. It is not 
straightforward to decide which signal level will be sufficient for efficient tracking.
If it is possible to operate with a signal of 6000 electrons, a fluence of
$\Phi\approx 3\times 10^{15}\,\Neq$ will be reachable. Tests with pixel detectors
irradiated to this level have been performed and show encouraging 
results \cite{ref:rohe-fi}.
This corresponds to the radiation level at a radius of 8\,cm, leaving only the 
radiation hardness of the
innermost (4\,cm) layer unsolved. Strategies to reach the level of radiation
hardness required there are the subject of the CERN-RD50 collaboration \cite{ref:rd50}.
When a pixel detector is operated at such a high fluence,
charge sharing will be much reduced due to
the decreasing Lorentz angle and the diminished range of signal collection depth. 
This will eventually limit the spatial resolution to the binary resolution of pitch/$\sqrt{12}$.
If the value of better than $20\,\umu$m has to be achieved also at fluences above 
$10^{15}\,\Neq$, the pixel pitch has to be reduced which requires
a redesign of the readout chip. In this case it has also to be reconsidered, whether the 
analogue signal processing performed by the present CMS pixel readout chip is still useful.

\subsection{Readout Chip}

The radiation hardness of ASICs fabricated in $0.25\,\umu$m technology 
seems to be sufficient up to ionisation doses\footnote{The tolerance against charged particles
at this high level has still to be checked. A reduction of the charge carrier
mobility might lead to speed limitations and the increase of the junction
leakage current could discharge dynamic nodes and lead to an increased
standby power.} exceeding 250\,kGy. The performance of the readout chip is 
therefore limited by the readout losses due to the high rate of tracks. As one would 
like to avoid a complete redesign of the present CMS pixel chip, the limits 
of its architecture will be discussed in section~\ref{sec:loss-slhc},
after the main mechanisms for data loss has been explained in the next paragraph.

\subsubsection{Architecture and Readout Losses of the CMS Pixel Readout Chip\label{sec:ro-losses}}

The architecture of the readout chip was adjusted to the environment of the
LHC and is described in detail in~\cite{ref:wolfram}. Here only the features necessary to
understand the inefficiency mechanisms are discussed.

Each pixel contains a preamplifier, shaper and comparator. When the amplitude of
the shaper exceeds the comparator threshold, the periphery of the chip, which is
shared by two columns of pixels, is notified and the amplitude is stored in
a sample-and-hold capacitor. The double column periphery then creates a
{\em time stamp} and initiates a scan (``column drain'') which 
copies the amplitude of all hit pixels into the data buffers. After the trigger latency
the buffers are either deleted or read out. In case of a trigger confirmation 
the double column is reset after the readout. During all these operations data 
can be lost by the mechanisms discussed below. The probability of their occurrence
was studied in detail  with Monte-Carlo simulations \cite{ref:danek-rate}. The numbers
given in the following are the expected inefficiencies for the 4\,cm layer at nominal 
(full) LHC luminosity.
\begin{description}
\item[\rm\em Pixel Busy:]
If a pixel is hit twice before the charge is transferred to the data buffer
the second hit is lost. Due to the fast draining mechanism and
the low pixel occupancy such events are unlikely ($0.21\,\%$).
\item[\rm\em Double Column Busy:]
During a column drain all empty pixels in the same double column are sensitive.
If a new pixel is hit during the scan, a time stamp is created and the column
drain is ``stacked'' until the previous one is completed. However, this stack 
is limited to two pending column drains and further hits will be lost until the
first of the column drains is completed. The probability for a
hit being lost by this mechanism is $0.25\,\%$.
\item[\rm\em Buffer Overflow:]
The number of time stamp (12) and data (32) buffers was adjusted to the
data rate for the 4\,cm layer and the buffer overflow rate is low
($0.17\,\%$ each).
\item[\rm\em Reset Loss:]
The dominant source of data loss is caused by the reset of the double column
after each triggered readout. The reset destroys the history of this double
column toghether with its data. If there are two events that pass the
level-1-trigger in a short time interval 
($\textrm{trigger latency of\ }3.2\,\umu\mrm{s} + \textrm{time of readout}$) and 
both events have hits in the same double column, then the hits belonging to 
the 2nd trigger (in this double column) are lost. This inefficiency becomes 
serious ($\approx 3\,\%$) if  the track rate (4-cm-layer) and the trigger 
rate (100\,kHz) are high. If the trigger latency increases,
this inefficiency will also increase.
\end{description}
\begin{figure}
\centering\includegraphics[width=1\linewidth]{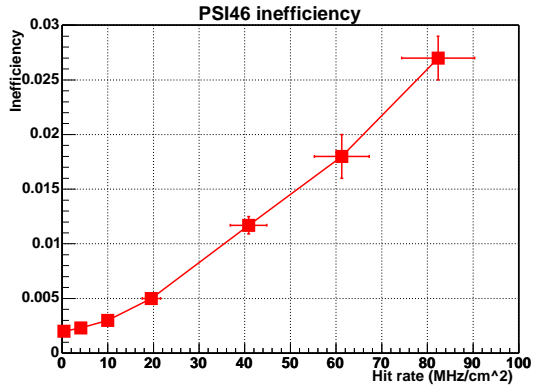}
\caption{Measured inefficiency of the CMS pixel readout chip as a function of the
    rate of perpendicular and uncorrelated tracks \cite{ref:wolfram}
    \label{fig:psi-testbeam}}
\end{figure}
All these inefficiencies add up to $3.8\,\%$ for the 4\,cm layer and about 
$1\,\%$ for the larger radii. 

The inefficiency of the
readout chip as a function of the track rate has been measured in a testbeam
experiment as shown in Fig.~\ref{fig:psi-testbeam} \cite{ref:wolfram}. As the
conditions there are slightly different from the final situation in CMS (impact
angle is $90^\circ$ and particles are uncorrelated) it is difficult
to relate this results to the LHC like conditions. However, the measured
losses approximately agree with the simulations explained above.

\subsubsection{Performance of the Present Readout Architecture at an upgraded LHC\label{sec:loss-slhc}}

With a further increase of the track rate when the luminosity is upgraded, the
readout losses will also increase. As not all incefficiency mechanisms have the same
rate dependence, a Monte-Carlo simulation was used to estimate the 
readout losses at a luminosity of $10^{35}\,$cm$^{-2}$s$^{-1}$. 
To keep the time stamp buffer overflow losses below 1\,\% the number of 
buffers has to be increased from 12 to about 60.
A similar scaling applies to the
data buffers. The use of a very compact $0.13\,\umu$m technology would 
certainly help to limit the increase of the peripheral chip area.

\begin{figure}
\centering\includegraphics[width=1\linewidth]{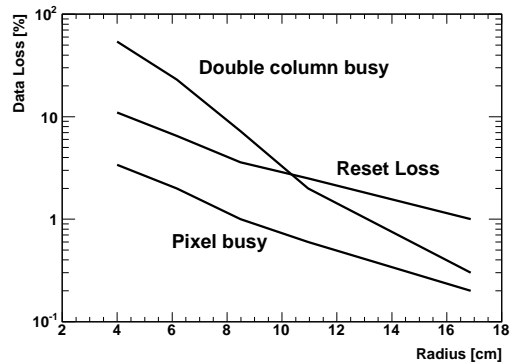}
\caption{Estimate of the radial dependence of the readout losses
      at a luminosity of $10^{35}\,$cm$^{-2}$s$^{-1}$  \label{fig:slhc-loss}}
\end{figure}
The dependency of the three remaining data loss mechanisms 
on the distance from the interaction point is
shown in Fig.~\ref{fig:slhc-loss}. With increasing particle rate (or decreasing
radius) the probability of a double column to be hit more than three times
during the time of a column drain increases dramatically. This limits the use of the present CMS-pixel readout at the tenfold LHC-luminosity to radii above 10\,cm.
For smaller radii a more elaborate redesign of the chip is necessary.

The reset loss does not increase as drastically with increasing particle
rate as it mainly depends on the trigger latency and rate which
are assumed to stay constant. The increase visible in Fig.~\ref{fig:slhc-loss}
is due to  the higher double column occupancy and the longer time necessary 
to read out the increased number of hits per triggered event.

The pixels are still small enough such that the probability to be hit twice
within a short time (pixel busy) remains relatively low.


\subsection{Cost Considerations}


\begin{table}
\caption{Rough cost estimate for one CMS pixel barrel module\label{tab:modul}}
\begin{center}
\begin{tabular}{lrr}
\hline
Component&\multicolumn{2}{c}{Costs [CHF] per}\\
&module&cm$^2$\\
\hline
16 readout chips ($0.25\,\umu$m)&400&40\\
1 Sensor (Si, n-in-n)&700&65\\
Bump bonding&1200&110\\
Fine-pitch hybrid (HDI)&300&30\\
Baseplate (Si$_3$N$_4$)&50&5\\
Cables and control chips&250&25\\
One Optical link&250&25\\
\hline
Sum& 3150&300\\
\hline
\end{tabular}
\end{center}
\end{table}

A rough cost estimate for a CMS pixel barrel module with an
sensitive area of $10.6\,\mrm{cm}^{2}$ is given in Tab.~\ref{tab:modul}. 
All the components added up lead to a price of roughly 300\,CHF/cm$^2$. 
This cost estimate ignores all off-module components like mechanics, cooling,
cabling (not directly mounted on the detector modules), 
readout and control modules (in most cases VME), and power supplies. Those
components contribute significantly to the total cost, but depend only weakly 
on the detector concept. If the present hybrid pixel technology is found to be the
only possible one at fluences above $10^{15}\,\Neq$, a configuration
of three pixel layers at radii of 8, 11 and 14\,cm is conceivable. They would
add up to a total area of about 1.1\,m$^2$, assuming a 53\,cm long barrel. 

\section{Pixel Detectors at intermediate Radii\label{sec:mid-radii}}

At radii larger than 15\,cm neither the track rate nor the radiation hardness 
represents a major limit for pixel detectors. As the area to be covered
increases with the radius, cost issues become more important.
The most effective items for a possible reduction are bump bonding 
and sensors (see Tab.~\ref{tab:modul}). 

The most cost driving requirement of the sensor's n-in-n technology is 
the need of double sided processing. As the radiation hardness required at radii between
15 and 25\,cm is still in the range of $5-10\times 10^{14}\,\Neq$, the collection
of electrons is more favourable. This naturally leads to n-in-p sensors, an
option currently investigated by the RD50 collaboration \cite{ref:rd50}.
It will require single sided processing only and might be available 
on 150\,mm wafers. This  offers the chance of getting three large 
modules (e.g. $32\times 80\,$mm$^2$) per wafer which might reduce the 
cost per area by more than a factor of three.

The disadvantage
of the single sided process is that due to the absence of the guard rings on the
back side, all sensor edges are at the potential of the applied bias voltage. 
As a consequence of the  considerable radiation damage, high voltages 
(400-600\,V) will be required to obtain a close to complete charge collection.
Precautions to protect the readout electronics from  destructive discharges will
have to be taken, like e.g. the introduction of a thin Kapton film between readout 
chips and sensor edge.

Bumpbonding is in principle well established and widely used in industry. The cost 
driving requirement of particle physics is the small pitch below $100\,\umu$m.  
If this number is relaxed to a value above $250\,\umu$m, a wide range of 
cheap industrial packaging processes will become available (e.g. the IBM-C4 
process \cite{ref:ibmc4}). The track density in this area will allow pixel areas
of the order of $100\,000\,\umu$m$^2$. In addition the number of chip placements can 
be reduced by increasing the size of the readout chip to the largest practical 
dimensions (e.g. $16\times 20\,$mm$^2$).

The third most expensive component are the readout chips. The one-to-one
coverage of the sensitive area with readout chips will be kept and savings
can only be expected if the price of this kind of electronics further decreases
with time.

The design of the detector modules has to be done under stringent
cost constrains (e.g. choosing a more standard hybrid). If a level
of roughly 100\,CHF/cm$^{2}$ is achieved, it will be possible to 
equip two layers, e.g. at 18 and 22\,cm, with such detectors.

\section{Pixel Detectors at Larger Radii\label{sec:large-radii}}

For the radial region between 25 and 60\,cm an even more drastic cost
reduction of pixel detectors is necessary which can only be reached
if the full coverage of the active area with readout electronics is given up.
This is possible because the suitable size of a sensing element is a few square millimeters
while a readout channel can be integrated into an area of about $0.02\,$mm$^2$. 

The shape of a sensing element (``mini strip'' or ``macro pixel'') could be 
$200\times 5000\,\umu$m$^2$  leading to about 10\,000 channels per sensor. 
They could be read out by a small number of pixel chips with a much 
smaller cell size. The routing between the sensor cells and the inputs of
the readout chips could be performed via a thick (about $40\,\umu$m) polyimide 
layer. An alternative would be the MCM-D technique \cite{ref:mcmd} which in addition
allows the integration of other components on a module. As the readout chips
can be placed completely inside the active area of the sensor the sensor edges
need not be kept on ground potential and a single sided sensor would be
possible. Due to the moderate requirements in radiation hardness in this
area ($\Phi < 5\times 10^{14}\,\Neq$) ``traditional'' and cheaply available
DC-coupled p-in-n sensors can be used. 

If it is possible to build such a detector for about 50\,CHF/cm$^{2}$,
it is a good candidate for layers at radii of e.g. 30, 40 and 50\,cm.

\section{Conclusions}

The main tasks of the CMS pixel detector are the measurement of displaced
vertices and pattern recognition. The challenges in the hostile environment 
of LHC are the high level of radiation and the high rate
of tracks. The requirements in particle detection efficiency and spatial
resolution in CMS can be satisfied, however sometimes with little headroom.

In case of an LHC luminosity upgrade the increase of the track density will
require the extension of pixel detectors to larger radii. Large areas, however, 
can only be equipped with pixel detectors whose cost is considerably reduced.
A pixel system of 9 layers seems quite feasible:

\begin{description}
\item[Innermost layer:] The requirements of the innermost pixel layer (roughly 4\,cm)
    cannot be fullfilled with present technologies. Possible solutions are
    investigated by the RD50 collaboration. In addition the readout electronics
    for this environment has to be developed soon.
\item[Small radii:] The region starting from 8\,cm can probably be equipped with present
    pixel modules with moderately modified readout chips. 
    However, only binary spatial resolution can be reached.
\item[Intermediate radii:] For radii between 15 and 25\,cm a less expensive 
    pixel system with single sided sensors and industrial bumpbonding is 
    proposed.
\item[Large radii:] Radii between 30 and 60\,cm could be equipped
    with macropixels or ministrips.
\end{description}

If it is possbile to reach the targeted cost reduction such pixel systems
will be very attractive for many other applications.

\section*{Acknowledgment}
The authors would like to thank Kurt Gabathuler for useful discussions
and the careful review of the manuscript.

\bibliographystyle{elsart-num}
\bibliography{rohe}

\end{document}